\documentstyle[aps,epsf,twocolumn,prl]{revtex}

\begin{document}
\draft

\twocolumn[\hsize\textwidth\columnwidth\hsize\csname @twocolumnfalse\endcsname
%%%%%%%%%%%%%%%%%%%%%%%%%%%%%%
\title{Dynamics of the Ising Spin-Glass Model in a
Transverse Field}

\author{M. J. Rozenberg$^1$ and D. R.  Grempel$^2$ }

\address{$^1$Institut Laue--Langevin, BP156, 38042 Grenoble, France \\ 
$^2$D\'epartement de Recherche Fondamentale sur
la Mati\`ere Condens\'ee \\
SPSMS, CEA-Grenoble, 17, rue des Martyrs, 38054 Grenoble
Cedex 9, France}

\date{\today}
\maketitle
\widetext
\begin{abstract}
\noindent
We use quantum Monte Carlo methods and various analytic approximations  to solve  the Ising spin-glass model in a transverse field in the disordered  
phase. We focus on the behavior of the frequency 
dependent susceptibility of the system above and below the critical field. 
 We establish the existence of an exact equivalence
 between this problem and the single-impurity Kondo model at the quantum critical point. Our predictions for
the long-time dynamics of the model are in good agreement with 
experimental results on $\rm LiHo_{0.167}Y_{0.833}F_4$.
%        1         2         3         4         5         6         7         8
%2345678901234567890123456789012345678901234567890123456789012345678901234567890

\end{abstract}


]

\narrowtext

The physics of frustrated quantum spin systems is a fascinating
and rapidly growing area of condensed matter physics\cite{review}. 
One of the most widely investigated
systems is the Ising spin glass in a transverse field, a model that 
combines simplicity and experimental accessibility\cite{wu1,wu2}.
The Hamiltonian of the model reads,

\begin{eqnarray}
H = -{1\over \sqrt{N}} \sum_{i<j} J_{ij} S^{z}_{i} S^{z}_{j}
-\Gamma \sum_{i} S^x_i,  
\label{hamil}
\end{eqnarray}
where $S^{\mu}_i$, $\mu=x,z$ are components of a three dimensional 
spin-1/2 operator at the $i-$th site of a fully-connected 
lattice of size $N$ and the first sum runs over all pairs of sites. The exchange interactions $J_{ij}$ are independent 
random variables with a gaussian 
distribution of zero 
mean and variance $J=\left< J_{ij}^2\right>^{1/2}$, and $\Gamma$ is an applied   magnetic 
field transverse to the easy-axis $z$.
For $\Gamma = 0$ Eq. \ref{hamil} is the classical 
Sherrington-Kirkpatrick spin-glass model that has a second order 
phase transition at $T_{\rm g}^0 = J/4$. When $\Gamma$ is 
finite quantum fluctuations compete with the
tendency of the system to develop spin-glass order. As a result
a boundary $\Gamma(T)$ appears in the $\Gamma-T$ plane between spin-glass (SG) and paramagnetic (PM) phases. 

The model of Eq.\ref{hamil} is relevant for
 the compound $\rm LiHo_{0.167}Y_{0.833}F_4$, a site-diluted
derivative of the
dipolar-coupled Ising ferromagnet $\rm LiHoF_4$ \cite{wu1}.
An external magnetic field $H_{\rm t}$   
perpendicular to the easy axis splits the doubly degenerate ground 
state of the $\rm Ho^{3+}$ ion. This splitting is 
proportional to $H_{\rm t}^2$ and plays the role of $\Gamma$ in Eq. \ref{hamil} \cite{wu1,hansen}. Experimentally,  
$\rm LiHo_{0.167}Y_{0.833}F_4$ is paramagnetic at all temperatures  above a critical field $H_{\rm t}^{\rm c}\approx 12$ kOe \cite{wu1,wu2}. Below $H_{\rm t}^{\rm c}$ and for $T\gtrsim 25$ mK the behavior of the non-linear susceptibility indicates a second order transition line between SG and PM phases that ends at $T_{\rm g}^0=135$ mK and $H_{\rm t}=0$ \cite{note1}. Investigation of the long-time dynamics of this system  above this line  
has revealed the existence of a fast crossover in the field  dependence of the absorption at very low frequencies\cite{wu1,wu2}. This crossover is characterized by a steep increase of $\chi''(\omega\to 0)$ across an 
almost flat line in the $\Gamma-T$ plane 
at $\Gamma\approx \Gamma_{\rm c}$ and up to $T\sim T_{\rm g}^0$.
 
While the phase diagram of the model has been studied theoretically 
using a variety of methods \cite{usadel,ishii,goldschmidt}, much less is known about its dynamics that has only been discussed for
large $\Gamma / J$ \cite{fedorov}, and near the quantum critical point\cite{huse}.

In this paper we use a recently developed quantum Monte Carlo method 
(QMC) \cite{us} to find the exact numerical  paramagnetic solutions of the model throughout the $\Gamma-T$ plane, and also obtain analytic
 expressions that we derive in various limiting cases. 
We find that the behavior of $\chi''(\omega)$ above and below the critical 
field is qualitatively different. For $\Gamma > \Gamma_{\rm c}$ the 
zero-temperature spectrum of magnetic 
excitations has a gap $\Delta$ \cite{huse} that vanishes as $\Gamma\to 
\Gamma_{\rm c}\approx 0.76 J$. At finite but low temperatures, $T<\Delta$, the gap edge develops a tail of exponentially small weight. On the other hand, for small $\Gamma$ and low $T$, we find a narrow feature around $\omega=0$ whose intensity decreases rapidly with increasing field or temperature as spectral  weight is transferred to higher frequencies. 
We further demonstrate that at the quantum 
critical point the problem can be exactly mapped to the single-impurity Kondo model. The low-energy properties of the system in the neighborhood of this point are  characterized by a new energy-scale, $T_{0} \approx 0.08 J$. 
At finite temperature there is a crossover between the regimes just mentioned  that is  essentially controlled 
by $\Gamma$ up to $T\sim T_{\rm g}^0$. We finally present detailed predictions for the $\Gamma$ and 
$T$-dependence of the low-frequency response that are in good agreement with 
the experimental results on 
${\rm LiHo_{0.167}Y_{0.833}F_4}$\cite{wu2}.

Bray and Moore \cite{bm} have shown that the quantum spin-glass
problem can be exactly transformed into a single-spin problem with 
a time-dependent self-interaction $Q(\tau)$ 
determined by the feedback effects of 
its coupling to the rest of the spins.
As we have shown elsewhere for a related problem\cite{us},  
much progress can be made by eliminating the 
self-interaction in favor of  
 an auxiliary fluctuating time-dependent field $\eta(\tau)$ 
coupled to the spins. The free-energy per site ${\cal F}$ in the paramagnetic 
phase can then be written as

\begin{eqnarray}
\label{free-energy}
\beta {\cal F} = && \min_{Q(\tau)}
\left\{
\frac{J^2}{4}\int_{0}^{\beta}\int_{0}^{\beta}d\tau d\tau' 
Q^{2}(\tau-\tau') -\ln Z_{\rm loc}[Q]\right\},
\nonumber\\
Z_{\rm loc}[Q]= && \int {\cal D}\eta \exp\!
\left[-{1 \over 2}\!\int_0^\beta\!\int_0^\beta\!d\tau d\tau'
Q^{-1}(\tau,\tau'){\eta}(\tau){\eta}(\tau')\right]
\nonumber\\
&&\times{\rm Tr {\cal T}}\exp \left[
\int_0^\beta\!d\tau \left[ J\eta(\tau){S^z}(\tau)
+\Gamma {S^x}(\tau) \right] \right],
\end{eqnarray}
where ${\cal T}$ is the time-ordering operator along the imaginary-time axis 
$0\le \tau \le \beta$. $Z_{\rm loc}$ can be thought of as the average partition function
of a spin in an effective magnetic field
$\vec{h}_{\rm eff}=J{\eta}(\tau) {\hat {\rm e}}_z + \Gamma {\hat {\rm e}}_x$
whose $z$-component is a random gaussian function with variance $Q(\tau)$. 
The latter is determined by functional minimization of (\ref{free-energy})
which gives the self-consistency condition \cite{bm} 

\begin{eqnarray}
\label{selfconsist}
Q(\tau) = \left< {\cal T}
{S^z}(\tau)  {S^z}(0)\right>_{\eta(\tau)},
\end{eqnarray}
where the average is taken with respect to the 
probability density associated to $Z_{\rm loc}$. 
We solved Eqs. \ref{free-energy}-\ref{selfconsist} iteratively using the QMC technique 
that we have described elsewhere \cite{us}. 
The imaginary time axis is discretized in up to $L=128$ 
time slices with $\Delta \tau = \beta /L \le 0.5$. 
An iteration consists of at least 20,000 QMC 
steps per time-slice and self-consistency is
generally attained after about eight iterations except very close
to the quantum critical point. We mapped the spin-glass transition line in 
the $\Gamma-T$ plane using the well known stability criterion\cite{bm}
$1= J \chi_{\rm loc}$ where $\chi_{\rm loc}=\int_0^\beta d\tau Q(\tau)$ 
is the local spin-susceptibility. 
We found a second-order transition line ending at a quantum critical point at 
$T=0$ in agreement with previous work \cite{usadel,ishii,goldschmidt}.
Going down in temperature to $T \sim 10^{-2}J$ and extrapolating the results to
$T=0$ 
 we determined a precise value for the critical field $\Gamma_{\rm c}/J = 0.76 \pm 0.01$, which lies in between previous estimates \cite{huse,goldschmidt}.

\begin{figure}
\epsfxsize=3.5in
\epsffile{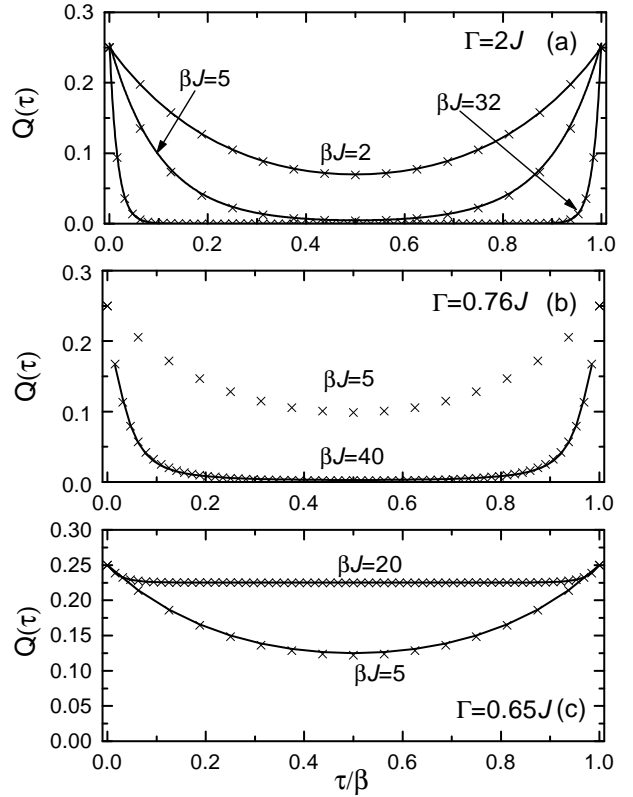}
\caption{$Q(\tau)$ as function of $\tau /\beta$. 
The crosses correspond to QMC data. The error bars are smaller than the size of the crosses. The solid lines are obtained using the analytic expressions discussed in the text.}
\label{fig1}
\end{figure}

We have studied the dynamical properties of the paramagnetic state throughout the $\Gamma-T$ plane, even below $T_{\rm g}$ where it is unstable. Indeed, the analysis of the evolution of the paramagnetic solution for small $\Gamma$ provides insight on the physics of this problem as the sates below and above $T_{\rm g}$ are continuously connected.
In Fig.\ref{fig1} we show the correlation function 
for several values of $\Gamma$ and $T$. 
For $\Gamma > \Gamma_{\rm c}$ (panel a), $Q(\tau)$  
decays exponentially with a time-constant 
$\tau_0\approx \Gamma^{-1} \ll \beta$ that depends only weakly on $T$. This behavior is characteristic of the existence of a gap $\Delta \sim \Gamma $ in the excitation spectrum of the system. For $\Gamma < \Gamma_{\rm c}$  and $T\ll J$ (panel c), $Q(\tau)$ also decays very rapidly for short times, 
$\tau \lesssim \tau_0 \sim \beta^{-1}$. 
For $\tau \gtrsim \tau_0$, however, it exhibits a very slow variation which indicates that the presence of excitations in the low-energy end 
of the spectrum,  $\omega \ll T$. 
With increasing temperature  $\tau_0$ increases and reaches a value ${\cal O}(J^{-1})$ when $T\sim J$. It then becomes no longer possible 
to distinguish two different time scales.
The case $\Gamma=\Gamma_{\rm c}$ at low temperature (panel b) is intermediate between the other two and the long-time behavior becomes a power-law, $Q(\tau)\propto \tau^{-2}$ as $T\to 0$, as anticipated by Miller and Huse \cite{huse} using internal consistency arguments. 
The solid lines in Fig. \ref{fig1} are the results of various analytic 
approximations that we discuss next. We begin by considering $\Gamma/J \gg 1$. In this case, the effective fields 
 appearing in Eq. \ref{free-energy}
are dominated by their $x-$component. We may thus evaluate the 
trace of the time-ordered exponential under the integral in $Z_{\rm loc}$ using a low-order cumulant expansion. 
To the lowest non-trivial order we find,
\begin{equation}
\chi''(\omega)=\frac{{\rm sgn}(\omega)}{2\Gamma  J m_x}
\sqrt{(2\Gamma J m_x)^{2}-(\omega^2-\Gamma^2)^{2}},
\label{highgamma}
\end{equation}
where $m_x=1/2 \tanh(\beta\Gamma/2)$ is the zero-th order 
transverse magnetization. Eq. \ref{highgamma} predicts the existence of  
a gap $\Delta = \sqrt{\Gamma^2 - 2\Gamma J m_{x}}$
in the excitation spectrum. This gap has a very weak temperature 
 dependence for $T\ll\Gamma$ and, at $T=0$, vanishes at $\Gamma_{\rm c}/J = 1$, overestimating the critical field. 
It can be shown \cite{us2} that, 
for $T \ll \Delta$, the gap-edge develops a tail that carries a small weight ${\cal O}(e^{-\Delta/T})$. 

This procedure breaks down at $\Gamma / J\sim 1$
 and a different 
approach must be taken in order to describe the physics at low fields. 
At $\Gamma = 0$, the problem reduces to the classical 
Sherrington-Kirkpatrick model and $Q(\tau)\equiv 1/4$ at all temperatures. 
We thus expect that for $\Gamma / J \ll 1$ the effective field 
$\vec{h}_{\rm eff}(\omega_n)$
will be dominated by its $\omega = 0$ component. Indeed, 
setting  for the moment $\eta(\omega)=0$ for $\omega \ne 0$ in 
Eq. \ref{selfconsist} and performing the functional average solely over 
static fields $\eta(\tau)\equiv \eta_{0}$, we find 
\begin{eqnarray}
Q(\omega_n)=\left<\frac{\beta J^2\eta_0^2/4}{\Gamma^2 +
J^2 \eta_0^2} \right>_{\eta_0} \delta_{n,0}
+ \left< \frac{\Gamma m(\eta_0)
\cos \theta_x}{\omega_n^2+ \Gamma^2 + J^2 \eta_0^2} 
\right>_{\eta_{0}}, 
\label{qsa}
\end{eqnarray}
where $m(\eta_0)=1/2 \tanh(\beta \sqrt{\Gamma^{2}
+J^2\eta_{0}^{2}}/2)$ and $\cos \theta_x=
\Gamma/\sqrt{\Gamma^{2}
+J^2\eta_{0}^{2}}$. The average is performed with respect to the probability distribution, 
${\cal P}(\eta_0) \propto \exp({-\beta \eta_0^2 /2\chi_{\rm loc}}) 
\cosh ({ \beta\sqrt{\Gamma^2 + J^2\eta_0^2}/2})$. Using Eqs. \ref{free-energy} and \ref{qsa} we can deduce the region of validity of this 
ansatz from the estimates   
$|\eta(\omega_n)/\eta(0)|\sim 32 \Gamma T^2 /J^3$ for $T \ll J$,  
and $\sim \Gamma /(2\pi nT)$ when
$T \gg J$. 
Within this approximation, the imaginary part of the
response on the real axis is given by,

\begin{eqnarray}
\frac{\chi''(\omega)}{\pi \omega} = 
\left< \frac{\beta J^2\eta_0^2/4}{\Gamma^2 +
J^2 \eta_0^2} \right>_{\eta_0} \delta(\omega) +
\nonumber \\
\frac{\Gamma^2}{2\omega^2}
\frac{\tanh(\beta |\omega|/2){\cal P} (\sqrt{\omega^2-\Gamma^2})}{\sqrt{\omega^2-\Gamma^2}}.
\label{qsa1}
\end{eqnarray}
In this regime the relaxation function $\chi''(\omega)/\omega$ splits into two 
contributions, 
an elastic peak at $\omega=0$ 
and a continuum starting at $\omega = \Gamma$.
The fraction of spectral weight contained in each of 
these two contributions
is determined by the transverse field and the temperature.
Evaluating the coefficient of the $\delta$-function in  Eq. \ref{qsa1} we find that the relative intensity of the central peak 
varies between $1- (8\Gamma T / J^2)^2$ for $T \ll J$, and
$J^2 / (4 \Gamma^2 + J^2)$ for $ T\gg J$.
The low-energy
 states represented by the $\delta$-function are responsible for the slow decay observed in the long-time behavior 
of $Q(\tau)$ at low $T$ in Fig.\ref{fig1}c. The fact that the central peak has zero width is a shortcoming of the approximation leading to Eq. \ref{qsa}
 as it does not capture 
the slow relaxational processes which broaden it \cite{us}. Nevertheless, the excellent agreement between 
the numerical and analytic results indicates that the width of the central peak must be, in any case, much smaller than the temperature. 
The high-energy states described by the inelastic part of the response control the exponential decay observed for short times. The decay rate predicted by 
Eq. \ref{qsa1} is $\tau_0^{-1}\sim J^2\beta/8$ for $T\ll J$, and 
$\tau_0^{-1}\sim J$ for $T\gg J$ in agreement with the numerical results.

The approximations discussed above 
can not be used near $\Gamma =\Gamma_{\rm c}$. 
However, one can still gain insight on the dynamics in the
critical region by exploiting
an interesting analogy between the model (\ref{hamil}) at the quantum critical
point and the single impurity Kondo problem that we establish next. 
We first perform a Trotter decomposition of the time ordered exponential 
in Eq. \ref{free-energy} and introduce intermediate states
$|\sigma \rangle \langle \sigma|$
at each imaginary time-slice $\tau$. The trace is now evaluated using
 the expression
\begin{equation}
\langle \sigma |e^{\Delta \tau 
\vec{h}_{\rm eff}(\tau) \cdot 
\vec{S}}|\sigma'\rangle \approx e^{\Delta \tau J\eta(\tau)\sigma}
\left(\delta_{\sigma \sigma'}+
\delta_{\sigma \bar{\sigma'}} \frac{\Gamma \Delta \tau}{2}+{\cal O}(\Delta\tau^2)\right),
\end{equation}
valid in the limit when the witdth of the time slice 
$\Delta \tau \to 0$. The partition function can then be expressed in terms 
of a sum over ``histories'', each of them defined
by a particular sequence of the eigenvalues $\sigma^z(\tau)=\pm 1/2$ of the intermediate states. Going over to the continuum limit and performing the gaussian integral
 over the auxiliary 
fields $\eta(\tau)$, we obtain

\begin{eqnarray}
Z_{\rm loc} = && \int {\cal D} \sigma^{z}
\exp \Big[
 \frac{J^2}{2} \int_0^{\beta}\int_0^\beta d\tau d\tau'
\sigma^z(\tau)Q(\tau - \tau')\sigma^z(\tau') 
\nonumber\\
&&  + \ \ln (\Gamma/2) \times {\text {(number of spin flips)}}
 \Big],
\label{yuval}
\end{eqnarray}
where by ``number of spin flips'' we mean the number of times that the
function $\sigma^z(\tau)$ changes sign in the interval
$0 \le \tau \le\beta$. This expression has a form analogous to that of the 
partition function
of the single-impurity Kondo model in the Anderson-Yuval 
formalism\cite{and-yuv}. In Eq. \ref{yuval}, $\Gamma$ and $J^{2}Q(\tau)$ play  
the roles of the
spin-flip coupling $J_{\pm}$ and the long-range Ising-like 
effective interaction in the Kondo model\cite{and-yuv}.  
Still, there is an important difference between the two problems. In the latter the Ising-like interaction is given, and behaves as 
$(2-\epsilon)/\tau^{2}$ with $\epsilon\propto J_{z}$ as $T\to 0$\cite{and-yuv}.  When the dynamics of the impurity-spin is controlled by the strong-coupling fixed point, its  time-dependent correlation function is also $\propto \tau^{-2}$ at long times. In contrast, in our problem, $Q(\tau)$ 
is {\it a priori} unknown. However, one realizes that, if for some $\Gamma$
 the asymptotic behavior of $Q(\tau)$ is $\sim \tau^{-2}$ then, by virtue of (\ref{yuval}) and the identifications just made, the two problems become equivalent at that value of the field and $T=0$. This is indeed the case  
at $\Gamma_{\rm c}$ 
where, as shown in Fig. \ref{fig1}b, the self-consistent solution of 
the problem at low $T$ can be accurately fitted 
by the finite-temperature generalization of Anderson and Yuval's Ising-like interaction\cite{and-yuv} using $\tau_0\sim \Gamma_{\rm c}^{-1}$ as a 
short-time cutoff. 
This analogy between our problem and the Kondo model provides a simple way to estimate the energy-scale associated with the low-energy excitations at the quantum critical point. It is well known\cite{krishnamurty} that the local susceptibility of the Kondo
 impurity is given by $T_0 \chi_{\rm loc}(T)=\phi(T/T_0)$, where $\phi$ is
a universal function with $\phi(0)\approx 0.0796$ \cite{krishnamurty} and $T_0$ is
the Kondo scale. Since in our case $\chi_{\rm loc}=J^{-1}$ at the quantum critical point, it follows that $T_0\approx 0.0796 J$. We expect the quantum critical region to extend up to a temperature $T_{\rm qc}$ of this order. Our detailed numerical results for the temperature dependence of the local susceptibility are consistent with $T_{\rm qc}\sim T_0/4$\cite{us2}.

Finally, we would like to discuss the experimentally observed crossover dynamics in $\rm LiHo_{0.167}Y_{0.833}F_4$ in the light of the results presented here.
 By measuring the response of the system at 1.5 Hz, the experiment\cite{wu2} is probing the intensity of the low-lying excitations represented by the 
$\delta$-function peak in Eq. \ref{qsa1}.

\begin{figure}
\epsfxsize=3.5in
\epsffile{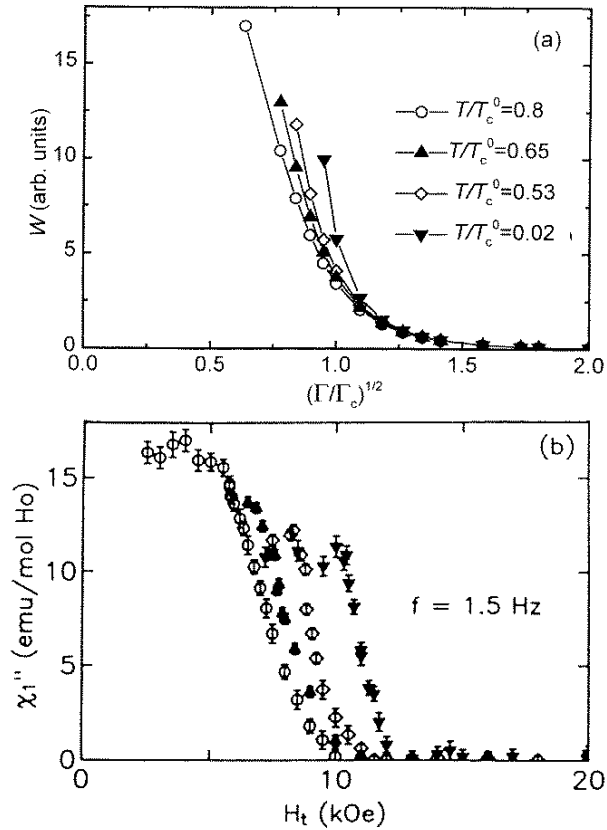}
\caption{(a) Spectral weight of the central peak in Eq. \protect{\ref{qsa1}} as 
a function
of $(\Gamma / \Gamma_{\rm c})^{1/2}$ for several temperatures.
(b) Imaginary part of the linear susceptibility of $\rm LiHo_{0.167}Y_{0.833}F_4$ at 1.5 Hz
 as a function of $H_t$. Data correspond to $T$=98, 82, 66 and 25 mK, from left to right. From Ref. \protect{\cite{wu2}}}
\label{fig2}
\end{figure}
\noindent In Fig. \ref{fig2} we compare the intensity of the latter (upper panel) and the experimental data of Ref. \cite{wu2} (lower panel) at various temperatures.  To make the connection with the experiment we plot the theoretical results as a function of $(\Gamma/\Gamma_{\rm c})^{1/2}$ since the splitting of the ground-state doublet of the Ho$^{3+}$ ion is proportional to the square of $H_{\rm t}$. 
The temperatures chosen for our  calculations are such that the ratios 
$T/T_{\rm g}^0$ correspond to the experimental values. There are no free parameters other than the overall scale of the $y$-axis. The theoretical curves end at the field $\Gamma_{\rm c}(T)$ where the condition $J\chi_{loc}=1$ is fulfilled.  For lower fields, the system enters the spin-glass phase and the experimental data become field-independent. As the figure demonstrates the overall behavior of 
the theoretical curves is in remarkable agreement with the experiment.
 Two important features are worth noticing. 
Firstly, the value of the field at the onset of absorption, 
$H_{\rm t}\approx H_{\rm t}^{\rm c}$ is almost temperature-independent which explains the flatness of the experimental crossover line.
Secondly, for $T\sim T_{\rm g}^0$, absorption starts well above $\Gamma_{\rm c}(T)$, meaning that precursor low-lying excitations appear in the paramagnetic phase long before the system freezes.

In conclusion, we have presented an exact numerical
solution of the infinite-range Ising spin-glass model in a  transverse
field in the paramagnetic phase. We worked out analytic approximations in different limiting cases that allow for a physical interpretation of the numerical data. In particular, we established for the first time an 
interesting connection between this problem and the single-impurity Kondo model.
Our prediction for dynamics of the model are in good agreement with 
experimental results on $\rm LiHo_{0.167}Y_{0.833}F_4$.


\begin{references}
\bibitem{note1}
There is some indication\cite{wu2} that the phase transition might become 
first-order below 25 mK, a  puzzling result \cite{mattsson,wu3} that 
we do not discuss in this paper.
\end{references}
\end{document}